\documentstyle[mprocl]{article}

\def\Journal#1#2#3#4{{#1} {\bf #2}, #3 (#4)}

\def\be{\begin{equation}}
\def\ee{\end{equation}}
\def\bea{\begin{eqnarray}}
\def\eea{\end{eqnarray}}
\begin{document}

%\title{NEWTONIAN LIMITS OF THE RELATIVISTIC COSMOLOGICAL PERTURBATIONS}
\title{RELATIVISTIC COSMOLOGICAL HYDRODYNAMICS}

\author{J. HWANG}

\address{Department of Astronomy and Atmospheric Sciences \\
         Kyungpook National University, Taegu, Korea}

\author{H. NOH}

\address{Korea Astronomy Observatory \\
         San 36-1, Whaam-dong, Yusung-gu, Taejon, Korea}


\maketitle\abstracts{
We investigate the relativistic cosmological hydrodynamic perturbations.
We present the general large scale solutions of the perturbation variables 
valid for the general sign of three space curvature ($K$), the cosmological
constant ($\Lambda$), and generally evolving background 
equation of state.
The large scale evolution is characterized by {\it a conserved gauge invariant
quantity} which is the same as a perturbed potential (or curvature) 
in the comoving gauge.}

%%%%%%%%%%%%%%%%%%%%%%%%%%%%%%%%%%%%%%%%%%%%%%%%%%%%%%%%%%%%%%
\section{Relativistic Cosmological Hydrodynamics}

%%%%%%%%%%%%%%%%%%%%%%%%%%%%%%%%%%%%%%%%%%%%%%%%%%%%%%%%%%%%%%
\subsection{Basic Equations}

The evolution of a homogeneous and isotropic model universe is described 
by the following equations (the third equation follows from the first two):
\bea
   H^2 = {8 \pi G \over 3} \mu - {K \over a^2} + {\Lambda \over 3}, \quad
       \dot \mu = - 3 H \left( \mu + p \right), \quad
       \dot H = - 4 \pi G ( \mu + p ) + {K \over a^2},
   \label{BG-eqs} 
\eea
where $H \equiv \dot a/a$; $\mu$ and $p$ are the energy density and 
the pressure, respectively.
A set of equations describing small but general scalar type perturbations 
of the above world model is the following:
\bea
   & & \dot \delta_v - 3 H {\rm w} \delta_v
       = - \left( 1 + {\rm w} \right) {k \over a} {k^2 - 3 K \over k^2} v_\chi
       - 2 H {k^2 - 3K \over a^2} {\sigma \over \mu},
   \label{P-1} \\
   & & \dot v_\chi + H v_\chi = - {k \over a} \varphi_\chi
       + {k \over a \left( 1 + {\rm w} \right)} 
       \left[ c_s^2 \delta_v + {e \over \mu}
       - 8 \pi G \left( 1 + {\rm w} \right) \sigma
       - {2 \over 3} {k^2 - 3 K \over a^2} {\sigma \over \mu} \right],
   \nonumber \\
   \label{P-2} \\
   & & {k^2 - 3 K \over a^2} \varphi_\chi = 4 \pi G \mu \delta_v,
   \label{P-3} \\
   & & \dot \varphi_\chi + H \varphi_\chi
       = - 4 \pi G \left( \mu + p \right) {a \over k} v_\chi - 8 \pi G H \sigma,
   \label{P-4}
\eea
where ${\rm w} \equiv p/\mu$ and $c_s^2 \equiv \dot p/\dot \mu$.
The fluid variables $\delta ({\bf k}, t)$, $v ({\bf k}, t)$,
$e ({\bf k}, t)$ and $\sigma ({\bf k}, t)$ are the relative density
perturbation ($\delta \mu /\mu$), the (frame independent) velocity, 
the entropic and anisotropic pressures, respectively.
The metric variables $\varphi ({\bf k}, t)$ and $\chi ({\bf k}, t)$ 
are the perturbed part of the three space curvature and the shear, respectively.
Eqs. (\ref{P-1}-\ref{P-4}) are written using the following gauge invariant 
combinations:
\bea
   \delta_v \equiv \delta + 3 (1 + {\rm w}) {aH \over k} v, \quad
       \varphi_\chi \equiv \varphi - H \chi, \quad
       v_\chi \equiv v - {k \over a} \chi.
   \label{GI-combinations}
\eea
$\delta_v$ is the same as $\delta$ in the comoving gauge ($v \equiv 0$), etc; 
we name the other gauge conditions as the uniform-curvature gauge 
($\varphi \equiv 0$), the zero-shear gauge ($\chi \equiv 0$),
the uniform-density gauge ($\delta \equiv 0$), etc.
$e$ and $\sigma$ are gauge invariant.

%%%%%%%%%%%%%%%%%%%%%%%%%%%%%%%%%%%%%%%%%%%%%%%%%%%%%%%%%%%%%%
\subsection{Closed Form Expressions}

Combining eqs. (\ref{P-1}-\ref{P-4}) we can derive a closed form expression
for the $\delta_v$ as
\bea
   & & \ddot \delta_v + \left( 2 + 3 c_s^2 - 6 {\rm w} \right) H \dot \delta_v
       + \Bigg[ c_s^2 {k^2 \over a^2} 
       - 4 \pi G \mu \left( 1 - 6 c_s^2 + 6 {\rm w} - 3 {\rm w}^2 \right)
   \nonumber \\
   & & \qquad \qquad \qquad
       + 12 \left( {\rm w} - c_s^2 \right) {K \over a^2}
       + \left( 3 c_s^2 - 5 {\rm w} \right) \Lambda \Bigg] \delta_v
   \nonumber \\
   & & \qquad
       = {1 + {\rm w} \over a^2 H} \left[ {H^2 \over a (\mu + p)} 
       \left( {a^3 \mu \over H} \delta_v \right)^\cdot \right]^\cdot
       + c_s^2 {k^2 \over a^2} \delta_v
   \nonumber \\
   & & \qquad
       = - {k^2 - 3 K \over a^2} { 1 \over \mu }
       \left\{ e + 2 H \dot \sigma
       + 2 \left[ - {1 \over 3} {k^2 \over a^2} + 2 \dot H
       + 3 \left( 1 + c_s^2 \right) H^2 \right] \sigma \right\}.
   \label{delta-eq} 
\eea
Notice that eq. (\ref{delta-eq}) is valid for general $K$, 
$\Lambda$, and $p = p(\mu)$.
The similar equation for $\varphi_\chi$ can be derived using eq. (\ref{P-3}).

%%%%%%%%%%%%%%%%%%%%%%%%%%%%%%%%%%%%%%%%%%%%%%%%%%%%%%%%%%%%%%
\subsection{General Solutions in the Large Scale}

On scales larger than the sound horizon (Jeans scale), 
ignoring the entropic and anisotropic pressures, 
eqs. (\ref{delta-eq},\ref{P-3}) lead to a general integral form solution as
\bea
   \varphi_\chi ({\bf k}, t)
   &=& 4 \pi G C ({\bf k}) {H \over a} \int_0^t {a (\mu + p) \over H^2} dt
       + {H \over a} d ({\bf k})
   \nonumber \\
   &=& C ({\bf k}) \left[ 1 - {H \over a} \int_0^t
       a \left( 1 - {K \over \dot a^2} \right) dt \right]
       + {H \over a} d ({\bf k}),
   \label{varphi_chi-sol}
\eea
where $C({\bf k})$ and $d({\bf k})$ are integration constants corresponding 
to the growing and decaying modes, respectively.
$\delta_v$ and $v_\chi$ follow from eqs. (\ref{P-3},\ref{P-4}), respectively, 
as:
\bea
   & & \delta_v ({\bf k}, t) = {k^2 - 3 K \over 4 \pi G \mu a^2} 
       \varphi_\chi ({\bf k}, t), 
   \label{delta_v-sol} \\
   & & v_\chi ({\bf k}, t)= - {k \over 4 \pi G ( \mu + p) a^2}
       \left\{ C ({\bf k}) \left[ {K \over \dot a} 
       - \dot H \int_0^t a \left( 1 - {K \over \dot a^2} \right) dt \right]
       + \dot H d ({\bf k}) \right\}.
   \nonumber \\
   \label{v_chi-sol}
\eea
{}For $K = 0 = \Lambda$ and ${\rm w} = {\rm constant}$, thus
$a \propto t^{2/[3(1+{\rm w})]}$, eqs. (\ref{varphi_chi-sol}-\ref{v_chi-sol})
become:
\bea
   \varphi_\chi ({\bf k}, t) 
   &=& {3 (1 + {\rm w}) \over 5 + 3 {\rm w}} C ({\bf k})
       + {2 \over 3 (1 + {\rm w})} {1 \over at} d ({\bf k})
   \nonumber \\
   &\propto& {\rm constant}, 
       \quad t^{- {5 + 3 {\rm w} \over 3 ( 1 + {\rm w} )} }
       \quad \propto \quad {\rm constant}, 
       \quad a^{- {5 + 3 {\rm w} \over 2} },
   \nonumber \\
   \delta_v ({\bf k}, t) 
   &\propto& t^{2(1 + 3 {\rm w}) \over 3 (1 + {\rm w})}, 
       \quad t^{- {1 - {\rm w} \over 1 + {\rm w} } }
       \quad \propto \quad a^{1 + 3 {\rm w}}, 
       \quad a^{- {3 \over 2} (1 - {\rm w}) },
   \nonumber \\
   v_\chi ({\bf k}, t)
   &\propto& t^{1 + 3 {\rm w} \over 3 (1 + {\rm w})}, 
       \quad t^{- {4 \over 3(1 + {\rm w}) } }
       \quad \propto \quad a^{1 + 3 {\rm w} \over 2}, 
       \quad a^{- 2}.
   \label{Sol-w=const}
\eea

%%%%%%%%%%%%%%%%%%%%%%%%%%%%%%%%%%%%%%%%%%%%%%%%%%%%%%%%%%%%%%%%%
\subsection{A Conserved Quantity}
 
$\varphi_v$ (the curvature fluctuation in the comoving gauge) is known to be 
conserved in the large scale limit independently of the changes in the 
background equation of state. 
Since we have $\varphi_v = \varphi_\chi - (aH / k) v_\chi$ from 
eq. (\ref{GI-combinations}), eqs. (\ref{varphi_chi-sol},\ref{v_chi-sol}) lead
to
\bea
   \varphi_v ({\bf k}, t) 
   &=& C ({\bf k}) \left\{ 1 + {K \over a^2}
       {1 \over 4 \pi G (\mu + p)} \left[ 1 
       - {H \over a} \int_0^t a \left( 1 - {K \over \dot a^2} \right) dt 
       \right] \right\}
   \nonumber \\
   & & + {K \over a^2} { H/a \over 4 \pi G ( \mu + p)} d ({\bf k}).
   \label{varphi_v-sol}
\eea
{}For $K = 0$ (but for a general $\Lambda$) we have
\bea
   \varphi_v ({\bf k}, t) = C ({\bf k}),
   \label{varphi_v-sol2}
\eea
with the {\it vanishing} decaying mode.
Thus, for $K = 0$, $\varphi_v$ is conserved for the generally time varying
equation of state, $p = p(\mu)$.
This conservation property of the curvature variable in a certain gauge
also applies to the models based on a minimally coupled scalar field or even on 
classes of generalized gravity theories \cite{GGT}.

%%%%%%%%%%%%%%%%%%%%%%%%%%%%%%%%%%%%%%%%%%%%%%%%%%%%%%%%%%%%%%%%%
\section{Newtonian Correspondences}

After a thorough investigation of the behavior of variables in the pressureless
limit in various gauge conditions we have identifed the following 
correspondences with the Newtonian perturbation variables which are valid
in a general scale:
\bea
   \delta_v \leftrightarrow \delta, \quad
       {k^2 - 3 K \over k^2} v_\chi \leftrightarrow \delta v, \quad
       - {k^2 - 3 K \over k^2} \varphi_\chi \leftrightarrow \delta \Phi,
   \label{Correspondences}
\eea
where $\delta$, $\delta v$, and $\delta \Phi$ in the right-hand-sides are 
the relative density fluctuation ($\delta \equiv \delta \varrho/\varrho$), 
the velocity fluctuation, and the potential fluctuation, 
in the Newtonian context, respectively.
Eqs. (\ref{P-1}-\ref{P-3}) can be compared with the continuity, 
the momentum conservation, and the Poisson's equations in the Newtonian
context, respectively.
Parts of these correspondences were studied in \cite{Lifshitz}.
A complete version of this work is presented in \cite{Newtonian}.

%%%%%%%%%%%%%%%%%%%%%%%%%%%%%%%%%%%%%%%%%%%%%%%%%%%%%%%%%%%%%%%%%
\section*{Acknowledgments}
This work was supported by the KOSEF, Grant No. 95-0702-04-01-3 and
through the SRC program of SNU-CTP.

%%%%%%%%%%%%%%%%%%%%%%%%%%%%%%%%%%%%%%%%%%%%%%%%%%%%%%%%%%%%%%%%%

\end{document}